\documentclass[12pt,reqno]{article}

\usepackage[latin1]{inputenc}
\usepackage[tbtags]{amsmath}
\usepackage{amsfonts}
\usepackage{amssymb}
\usepackage{amsxtra}
\usepackage{color}
\usepackage{hyperref}
\hypersetup{linktocpage=true}
\usepackage[all]{xy}

\def\be {\begin{equation}}
\def\ee {\end{equation}}
\def\bs#1\es{\begin{split}#1\end{split}}
\def\ba#1\ea{\begin{align}#1\end{align}}
\def\bg#1\eg{\begin{gathered}#1\end{gathered}}
\def\bea{\begin{eqnarray}}
\def\eea{\end{eqnarray}}

\def\ie{{\it i.e.}\ }
\def\eg{{\it e.g.}\ }

\def\a{\alpha}
\def\b{\beta}
\def\c{\chi}
\def\d{\delta}
\def\e{\epsilon}

\def\g{\gamma}

\def\h{\eta}
\def\i{\iota}
\def\k{\kappa}
\def\l{\lambda}

\def\m{\mu}

\def\o{\omega}
\def\O{\Omega}
\def\p{\psi}

\def\q{\theta}

\def\s{\sigma}

\def\x{\xi}

\def\pa{\partial}
\def\na{\nabla}
\def\fr{\frac}
\def\sq{\sqrt}

\def\tr{\text{tr}}

\def\bls{\bigg [}
\def\brs{\bigg ]}

\def\cL{\mathcal{L}}

\def\cA{{{\cal A}}}
\def\cM{\mathcal{M}}

\def\cB{\mathcal{B}}
\def\cV{\mathcal{V}}
\def\id{\rlap 1\mkern4mu{\rm l}}
\def\nn{\nonumber}
\def\lra{\leftrightarrow}
\def\ra{\rightarrow}

\def\dc{d^{(\i)}}
\def\pac{\partial^{(\i)}}

\def\wh{\widehat}

\def\ztwo{\mathbb{Z}_2}
\renewcommand{\bar}{\overline}

\newcommand{\til}[1]{{\tilde {#1}}}
 
\newcommand{\tb}[1]{{\text{\tiny {(#1)}}}}
\newcommand{\hoch}[1]{${}^{#1}$}

\numberwithin{equation}{section}
\numberwithin{figure}{section}

\title{\vspace{-1cm}
\begin{flushright}\normalsize
\hfill{KCL-MTH-11-09}\\
\hfill{Imperial/TP/11/KSS/01}\\
\hfill{NSF-KITP-11-054}
\end{flushright}
\vspace{1.5cm}
\bf Ectoplasm with an Edge}
\date{\vspace{-2cm}}

\newcommand{\imperial}{{\it Theoretical Physics Group, Imperial College London, Prince Consort Road, London SW7 2AZ, UK}}
\newcommand{\kings}{{\it Department of Mathematics, King's College, University of London, Strand, London WC2R 2LS, UK}}
\newcommand{\kitp}{{\it Kavli Institute for Theoretical Physics, University of California, Santa Barbara CA 93106, USA}}
\newcommand{\auth}{
P.S. Howe\,\footnote{email: paul.howe@kcl.ac.uk}\hoch{\star}, T.G. Pugh\,\footnote{email: thomas.pugh08@imperial.ac.uk}\hoch{\dagger}, K.S. Stelle\,\footnote{email: k.stelle@imperial.ac.uk}\hoch{\dagger\ddagger} and C. Strickland-Constable\,\footnote{email: charles.strickland-constable08@imperial.ac.uk}\hoch{\dagger}}

\begin{document}

\setcounter{page}{0}

\maketitle

\thispagestyle{empty}

\auth

${}^\star$ \kings

${}^\dagger$ \imperial

${}^\ddagger$ \kitp

\begin{abstract}
\normalsize
The construction of supersymmetric invariant actions on a spacetime manifold with a boundary is carried out using the ``ectoplasm'' formalism for the construction of closed forms in superspace. Non-trivial actions are obtained from the pull-backs to the bosonic bodies of closed but non-exact forms in superspace; finding supersymmetric invariants thus becomes a cohomology problem. For a spacetime with a boundary, the appropriate mathematical language changes to relative cohomology, which we use to give a general formulation of off-shell supersymmetric invariants in the presence of boundaries. We also relate this construction to the superembedding formalism for the construction of brane actions, and we give examples with bulk spacetimes of dimension 3, 4 and 5. The closed superform in the 5D example needs to be constructed as a Chern-Simons type of invariant, obtained from a closed 6-form displaying Weil triviality.
\end{abstract}

\newpage

\noindent\rule\textwidth{.1pt}		
\tableofcontents
\vspace{20pt}
\noindent\rule\textwidth{.1pt}

\setcounter{page}{1}
\setcounter{footnote}{0}

\section{Introduction}

The representation of supersymmetry on spacetimes with boundaries has become an important issue owing to the advent of AdS/CFT dualities and the potential importance of Ho\v{r}ava-Witten \cite{Horava:1995qa,Horava:1996ma,Moss:2004ck,Moss:2008ng} or single-boundary constructions in cosmological \cite{Lukas:1998yy,Lukas:1998tt,Randall:1999ee,Randall:1999vf,Khoury:2001wf,Lehners:2008vx} and particle-physics \cite{Randall:1999ee,Pugh:2010ii} contexts. Arranging  for a heterogeneous system of a bulk theory and a boundary theory to be jointly supersymmetric was originally achieved by a traditional perturbative Noether construction. Boundaries can be treated in an ``upstairs picture'' with the boundary realised via a doubled spacetime endowed with a $\ztwo$ reflection symmetry. Corrections to bulk-field symmetry transformations and constraints then typically involve Dirac delta functions, which can pose problems at higher orders when they begin to pile up nonlinearly \cite{Horava:1995qa,Horava:1996ma}. Another way to treat boundaries is in the ``downstairs picture'' where boundaries are treated just as boundaries, and instead of delta functions, one deals with systems of boundary conditions for bulk fields in interaction with fields defined only on the boundary \cite{Moss:2004ck,Moss:2008ng}. In either formalism, the iterative construction is rather laborious and can be rather tricky, especially when the requirements of anomaly cancellation involve dealing with ``classical'' systems that have supersymmetry violation linked to gauge non-invariance by systems of Wess-Zumino consistency conditions \cite{Pugh:2010ii}. 

It would clearly be advantageous to have a formalism that automates the bulk-plus-boundary construction in a fashion similar to the standard tensor calculus of supersymmetry and supergravity \cite{Wess:1974jb,Ferrara:1978jt,Stelle:1978yr,Ferrara:1978wj,Stelle:1978wj}. This is precisely the aim of the ``supersymmetry without boundary conditions'' formalism which has recently been introduced \cite{Belyaev:2007bg,Belyaev:2008xk,Belyaev:2008ex} and put to use in \cite{Grumiller:2009dx,Belyaev:2010as}. This formalism relies upon the existence of an off-shell supersymmetry formalism, however, which is not available for all supersymmetric theories, including the key maximal super Yang-Mills and maximal supergravity theories.

In the study of invariants for such theories, another approach to the study of supersymmetric invariants has been developed: the ``ectoplasm'' formalism\footnote{The ectoplasm formalism, referring to out-of-body material, employs the use of  closed forms in full superspace and not just in the ``body'' or purely bosonic subspace.} \cite{Voronov,Gates:1997kr,Gates:1997ag}, which can be employed to codify the integrands of supersymmetric invariants through the study of closed forms in superspace (see, for example, \cite{Berkovits:2008qw,Bossard:2009mn,Bossard:2010bd,Bossard:2010pk}).  It is clearly of interest to relate the ``supersymmetry without boundary conditions'' formalism to the construction of supersymmetric invariants via closed forms in superspace. Relating these two approaches to supersymmetric theories on spaces with boundaries is the main aim of the present article.

The study of closed forms in superspace is a problem in superspace cohomology \cite{Berkovits:2008qw}. Studies of $p$-brane worldvolume theories, another context for supersymmetric boundary theory investigations, have pointed out the r\^ole played by relative cohomology in such contexts \cite{Kalkkinen:2002tk}. As we shall see, relative cohomology is precisely the mathematical language needed for the formulation of supersymmetry boundary theory problems in terms of closed forms and ectoplasm.

In this paper, we first review in Section \ref{Superfields with boundaries} the construction of supersymmetric theories on manifolds with boundaries, then in Section \ref{Superforms without boundaries} we review the construction of supersymmetric invariants via closed forms in superspace. In Section \ref{Superforms with boundaries}, we re-express the construction in terms of relative cohomology in superspace and establish the relation between the present formulation and the superembedding formalism \cite{Bandos:1995zw,Howe:1996mx,Howe:1997wf,Sorokin:1999jx}, which has been applied to supersymmetric brane worldvolume actions in \cite{Bandos:1995dw,Howe:1998tsa}.\footnote{The superembedding formalism was first proposed in the context of superparticles \cite{Sorokin:1988nj,Sorokin:1989zi}.} We also introduce the notion of a ``superboundary'' with half-projected surviving supersymmetry on the boundary. In Section \ref{examples}, we illustrate the construction with 3D, 4D and 5D examples. The 5D example involves the construction of a Chern-Simons type closed superform via the mechanism of Weil triviality \cite{Bonora:1986xd}.  In the Conclusion, we comment on some open issues and in Appendices \ref{conventions} and \ref{closure conditions} we  summarise our conventions and give details of the closure conditions for the various examples discussed in the text.

\section{Superinvariants with boundaries}
\label{Superfields with boundaries}
We begin by briefly reviewing the construction of a supersymmetric action on a manifold with a boundary.  The inclusion of the boundary has an immediate effect of breaking the diffeomorphism invariance under transformations that would move the boundary. This implies that surviving diffeomorphisms will be generated by a vector $\x^m$ such that $n_m \x^m = 0$, where $n_m$ is the outward pointing unit normal to the boundary, satisfying $n_m n^m = 1$. However, the commutator of two supersymmetry transformations generates a diffeomorphism given by
\ba
[ \d_{\e_1}, \d_{\e_2} ]   = \d_{\tilde \x} + \ldots\  , 
\ea
where $\tilde \x^m \sim \bar \e_{1} \g^m \e_{2} -  \bar \e_{2} \g^m \e_{1}$. We therefore need to also impose the condition that $n_m \tilde \x^m = 0$ in order to prevent repeated supersymmetry transformations from generating a diffeomorphism that would move the boundary and break the symmetry. This is solved by imposing the conditions
\ba
n_m \x^m & = 0\ ,  & n_m \g^m \e = \e\  . 
\label{xeConstraints}
\ea 
With these conditions in mind, we can attempt to construct actions that are invariant under the surviving supersymmetry. Usually, the Lagrangian for an invariant action is considered to be one that varies into a total derivative,
\ba
\d \cL =  \pa_{m} ( \bar \e \l^m +\bar \l^m \e  )\ , 
\ea
where $\l^m$ is some function of the fundamental fields in the theory. This means that if we integrate this Lagrangian over a manifold $\cM_0$ without boundaries, then we form an action 
\ba
S = \int_{\cM_0} d^d x \cL\ , 
\label{defSbasic}
\ea
which is invariant. However, if we now consider $\cM_0$ to have a boundary $\pa \cM_0$, then under supersymmetry this action will vary into
\ba
\d S  = \int_{\cM_0} d^d x \pa_{m} ( \bar \e \l^m +\bar \l^m \e  ) = \int_{\pa \cM_0} d^{d-1} x n_{m} ( \bar \e \l^m +\bar \l^m \e  ) \ . 
\label{varS}  
\ea
The traditional method for considering supersymmetric actions with boundaries was to impose some boundary conditions such as $ \fr12 (1- n_a \g^a ) \l^m n_m \big |_{\pa \cM_0}  = 0$ which would then force  \eqref{varS} to vanish subject to \eqref{xeConstraints}. This means that the action \eqref{defSbasic} is only invariant under supersymmetry up to boundary conditions and so we refer to this approach as being a ``SUSY with b.c.'' formalism. However, an alternative approach to these constructions has recently been developed \cite{Belyaev:2007bg} in which one instead modifies the action by the addition of a boundary localised term such that 
\ba
S = \int_{\cM_0} d^d x \cL + \int_{\pa \cM_0} d^d x \cL_B\  , 
\label{defSmodbasic}
\ea
where $\cL_B$ varies under supersymmetry as
\ba
\d \cL_B = - n_{m} ( \bar \e \l^m +\bar \l^m \e  )\ . 
\ea
This implies that the modified action \eqref{defSmodbasic} is invariant under supersymmetry without having to impose boundary conditions and accordingly this is known as the ``SUSY without b.c.'' formalism.  

To demonstrate explicitly how this formalism works, it is helpful to consider some examples. Firstly, we consider the case of 3D N=1 rigid supersymmetry. Here, actions are determined by an unconstrained superfield $J_0$ with components
\ba
J_0 | & = A\ , & D_\a J_0 | &  = \fr1{\sq2} i \c_\a\ ,  & D^\a D_\a J_0 | & = -  i F\ ,  
\ea
where $\big |$ denotes evaluation on the surface where all fermionic coordinates are set to zero. These component fields transform under supersymmetry as 
\ba
 \d A &= i \e^\a \c_\a\ ,  & \d \c_\a &=   \g^m{}_\a{}^\b \e_\b \pa_m A +   F \e_\a\ , & \d F  & = i  \e^\a \g^m{}_\a{}^\b \pa_m \c_\b\ , 
\label{var3DJ} 
\ea
where the 3D spinors are Majorana. 

The standard rule for constructing an F-density from this superfield is to build an action given by
\ba
 S = \int_{\cM_0} d^3 x\, i D^\a D_\a J_0 \big | =   \int_{\cM_0} d^3 x F\ .
\ea
Since this action varies into a surface term on $\pa \cM_0$, the traditional prescription for creating a supersymmetric action would be to impose boundary conditions on the component fields in $J_0$ that set this surface term to zero. However, in the ``SUSY without b.c.'' formalism, one modifies instead the F-density rule such that 
\ba
 S = \int_{\cM_0} d^3 x\, i  D^\a D_\a J_0 \big |  + \int_{\pa \cM_0} d^2 x J_0 \big |  =    \int_{\cM_0} d^3 x F + \int_{\pa \cM_0} d^2 x A\ . 
\ea
Using \eqref{var3DJ}, one then finds that the modified action varies into 
\ba
\d S =  \int_{\pa \cM_0} d^2 x\,  i \c^\a (  \e_\a - n_a \g^a{}_\a{}^\b \e_\b)\  , 
\ea
which is set to zero by the conditions \eqref{xeConstraints} regardless of the choice of boundary conditions on the component fields in $J_0$. 

As a second example, we review briefly the construction of ``SUSY without b.c.'' actions in rigid 4D N=1. Here, the action is determined by a chiral superfield $ J_0$ satisfying $\bar\na_{\dot \a} J_0 = 0 $ with components
\ba
J_0 | & = A\ , & D_\a J_0 | &  =  i \c_\a\ ,  & D^\a D_\a J_0 | & = - 2 i F\ . 
\ea
These components transform under SUSY as 
\ba
\d A &= i \e^\a \c_\a\ ,  & \d \c_\a &=   \s^m{}_\a{}^{\dot \b} \bar \e_{\dot \b} \pa_m A + F \e_\a\ ,  & \d F & = i  \bar \e^{\dot \a} \s^m{}_{\dot \a} {}^\b \pa_m \c_\b\ .
\label{var4DJ} 
\ea
The modified F-density rule in this case gives the action as
\ba
S &= \int_{\cM_0} d^4 x  \fr12 i \bls D^\a D_\a J_0 \big | + \bar D^{\dot \a} \bar D_{\dot \a} \bar J_0 \big |\brs   + \int_{\pa \cM_0} d^3 x \bls J_0 \big |  + \bar J_0 \big | \brs \nn \\
& =  \int_{\cM_0} d^4 x  ( F + \bar F) + \int_{\pa \cM_0} d^3 x  (  A  + \bar A)\ .
\ea
As before, it is easy to show that the variation of this modified F-density under \eqref{var4DJ} vanishes subject to a chirality condition on $\e$ such as \eqref{xeConstraints} without the need for any boundary conditions on $J_0$. 
  
\section{Superform invariants without boundaries}
\label{Superforms without boundaries}

We next review the standard superform approach to the construction of supersymmetric invariants ignoring boundary effects, following the discussion in \cite{Berkovits:2008qw}. Here we will show that the construction of an invariant action amounts to finding a super $d$-form $J_d$ that is closed, $d J_d = 0 $, and that is nontrivial under the cohomology equivalence $J_d \sim J_d + d\l_{d-1} $.

Consider a supermanifold $\cM$ with $d$ bosonic dimensions and $n$ fermionic dimensions. Let $\cM_0$ be the $d$-dimensional body of $\cM$, without boundary: $\pa \cM_0 = 0$. Let $b:\cM \ra \cM_0$ be the projection of the supermanifold onto its body and let $s: \cM_0 \ra \cM$  be a section of this projection.  Finally, let $J_d$ be an arbitrary super $d$-form on $\cM$. We can then consider the integral of the pullback of $J_d$ to $\cM_0$,
\ba
S = \int_{s(\cM_0)} J_d = \int_{\cM_0} s^* J_d\ ,
\label{defS}
\ea
which will form our action. As $\cM_0$ has no boundary, we find that $S[J_d] = S[J_d + d \l]$ so the action depends only on the de Rham cohomology class of $J_d$, \ie
\ba
J_d \sim J_d + d\l_{d-1}\  .
\label{cohomology1}
\ea
In order for the action to be nonzero, we require that $J_d$ not be exact. Since we consider only forms $\l_{d-1}$ that can be constructed from the physical fields of the theory, the cohomology can be non-trivial even if the spacetime has trivial topology. 

The condition that $S$ is invariant under supersymmetry is equivalent to the statement that $S$ is independent of the choice of even submanifold $s(\cM_0)\subset\cM$ and so is independent of the section $s$ chosen. To find the corresponding condition on $J_d$, we consider a one-parameter family of diffeomorphisms $f_t : \cM \ra \cM$. These give rise to a one-parameter family of even submanifolds $s_t (\cM_0) $ where $s_t = f_t \circ s$. The diffeomorphism $f_t$ then transforms $S$ to 
\ba
S_t = \int_{s_t (\cM_0) } J_d= \int_{\cM_0} s^* \circ f^*_t J_d\ . 
\ea
If $S$ is independent  of the section $s$, then it will be invariant under this diffeomorphism, so
\ba
\fr{d S_t}{d t }    = 0\  , 
\ea
for any diffeomorphism family $f_t$. Then, since
\ba
\fr{ d ( f_t^* \o )}{d t}  = \cL_v \o\ , 
\ea
for any superform $\o$, where $v$ is the vector field generating the diffeomorphism family $f_t$, we find that $S$ will be independent of the section $s$ if 
\ba
0 & = \fr{d S_t}{d t }\ , \nn \\
& = \int_{\cM_0} s^* \cL_v J_d\ ,  \nn \\
& = \int_{\cM_0} s^* ( d i_v J_d + i_v d J_d )\ , \nn \\
& = \int_{\cM_0} ( d ( s^* i_v J_d ) + s^* i_v d J_d )\ ,\nn \\
& = \int_{\cM_0} s^* i_v d J_d\ , 
\ea
which is solved for arbitrary $v$ if $ d J_d = 0 $. 

Thus we find that we can build a supersymmetric action from a super $d$-form $J_d$ with $ d J_d = 0 $ but where $J_d$ is cohomologically non-trivial under $J_d \sim J_d + d\l_{d-1} $. This implies that $J_d$ is a representative of a class in the de Rham cohomology group $H^d(\cM)$. As we have just seen that the action is independent of the section $s$ chosen, it is then natural to chose $s$ to be the section where all  fermionic coordinates vanish. 

To make the discussion more specific, we now need to refine the notation. We will label directions on the  supermanifold tangent space with an index $A = 1, \ldots , d+n$; the tangent space then splits into bosonic directions carrying an index $a = 0, \ldots, d-1$ and fermionic directions carrying an index $\a = 1, \ldots, n$. Any super $P$-form can be written as a sum of super $(p,q)$-forms with $p+q=P$\,:
\ba
\O_P = \fr{1}{P!} E^{A_P} \ldots E^{A_1}  \O_{A_1 \ldots A_P}  =  \sum_{i= 0}^P \o_{P-i,  i}\  , 
\ea
where 
\ba
\o_{p,q} = \fr{1}{p! q!}E^{\a_q} \ldots E^{\a_1} E^{a_p} \ldots E^{a_1}  \o_{a_1 \ldots a_p \a_1 \ldots \a_q } \  . 
\ea

Letting $P=d$, we can write the closure condition on $J_d$ as 
\ba
\na_{[A_1} J_{A_2 \ldots A_{d+1} )}  + \fr{d}{2} T_{[A_1 A_2 |}{}^{B} J_{ B | A_3 \ldots A_{d+1} ) }  = 0\ ,  
\label{dClosed}
\ea
where $\na_A$ is the covariant derivative on the supermanifold, $T_{AB}{}^C$ is the torsion and $[\ldots )$ indicates graded antisymmetrization. Similarly, we find that under the cohomology relation $J_d \sim J_d + d \l_{d-1}$, $J_d$ is equivalent to $J_d +\d J_d$ where $\d J_d$ is given by 
\ba
\d J_{ A_1 \ldots A_D } = \fr{1}{(d-1)!} \big(  \na_{[A_1} \l_{A_2 \ldots A_d)} + \fr{d-1}{2}  T_{[A_1 A_2 |}{}^{B} \l_{B | A_3 \ldots A_d )} \big) \  . 
\ea
We can then split the super $d$-form $J_d$ up into its super $(p,q)$-form parts $J_{p,q}$ with $p+q=d$ and can consequently analyse the constraint that $J_d$ be closed but not exact in terms of constraints on these parts \cite{Berkovits:2008qw}. We will refer to the nonvanishing $J_{p,q}$ with highest $q$ as the generator of $J_d$. This must satisfy 
\ba
T_{ (\a_1 \a_2|}{}^{a_p} J_{a_1 \ldots a_p | \a_3 \ldots \a_{q+2} )  }  = 0\ . 
\label{t0Closed}
\ea
In the case of rigid 3D N=1 supersymmetry, the only nonvanishing component of the torsion is $T_{\a \b}{}^c = i \g^c{}_{\a \b}$. Then \eqref{t0Closed} implies that the generator is of the form $J_{1,2} \sim \g_{1,2} J_0$ where $J_0$ is a superscalar and $\g_{1,2}$ is a single gamma matrix expressed as a superform.  Using the closure condition \eqref{dClosed}, we can then find the other parts of $J_d$ iteratively, obtaining
\ba
J_{\a \b \g} & = 0\ ,& 
J_{ a \a \b } &=   -  i \g_{a  \, \a \b} J_0\ ,\nn \\
J_{a b \a} & =   \e_{a b c} \g^{c}{}_{\a}{}^\b D_{\b} J_0\ ,&   
J_{a b c} & =   i  \e_{a b c} D^{\a} D_{\a} J_{0}\ .
\label{3DJ} 
 \ea
Similarly, in rigid 4D N=1 supersymmetry, the nonvanishing components of the torsion are just ($T_{\a \dot \b}{}^c =  i  \s^c{}_{\a \dot \b}$\,,  $T_{\dot \a \b}{}^{c} = i  \bar \s^{c}{}_{\dot \a \b} $). This means that \eqref{t0Closed} implies that the generator of $J_d$ is given by $J_{2,2} \sim \g_{2,2} J_0$. Once again, we can use \eqref{dClosed} iteratively to construct the closed superform $J_d$\,, which is given by  \cite{Gates:1997ag}
\ba
J_{a b \a \b } & =    2  \s_{ a b \, \a \b} \bar{J}_0 \ , &
J_{a b \dot \a \dot \b} & =    2  \bar \s_{ a b \, \dot \a \dot \b} {J}_0\ , \nn \\
J_{a b c \a} &=   \e_{a b c d} \s^d{}_{\a \dot \a} \bar D^{\dot \a} \bar J_0\ , & 
J_{a b c \dot \a} &=  \e_{a b c d } \bar \s^d{}_{\dot \a  \a} D^{\a}  J_0\ ,\nn \\
J_{a b c d} & =  - i \e_{a b c d} \fr12 \big(  D^\a D_\a  J_0  +  \bar D^{\dot \a} \bar D_{\dot \a} \bar J_0 \big)\ ,&
\bar D_{\dot \a} J_0 &= 0\ ,
\label{4DJ}
\ea 
with all other components vanishing. The unmodified F-density rules are then obtained by substituting these closed superforms into \eqref{defS}.

\section{Superform invariants with boundaries}
\label{Superforms with boundaries}

We now combine the ideas of Sections \ref{Superfields with boundaries} and  \ref{Superforms without boundaries}. This will allow us to arrive at a prescription for deriving the boundary modifications to the F-density rules, resulting in a ``SUSY without b.c.'' action. To do this, we again begin by considering a supermanifold $\cM$ with $d$ bosonic dimensions and $n$ fermionic dimensions, a $d$-dimensional body $\cM_0$ and a projection to the body $b : \cM \ra \cM_0$ with a section $s : \cM_0 \ra \cM$ . However, we now consider $\cM_0$ to have a boundary $\pa \cM_0 $ and a mapping $c : \pa \cM_0 \ra \cM_0$. As before, let $J_d$ be an arbitrary $d$-form on $\cM$ but also let $I_{d-1}$ be an additional arbitrary $(d-1)$--form on $\cM$. We again consider building an action by integrating over $\cM_0$ but now, motivated by the ``SUSY without b.c.'' approach, we also include an additional boundary-localised part integrated over $\pa \cM_0$\,:
\ba
S = \int_{\cM_0} s^* J_d \; - \int_{\pa \cM_0} c^* \circ s^* I_{d-1}\  . 
\label{defI}
\ea
The equivalence \eqref{cohomology1} now becomes modified to
\ba
\label{relcohomology1}
\begin{pmatrix}
J_d  \\
I_{d-1}
\end{pmatrix} \sim \begin{pmatrix} 
J_d + d\l_{d-1}  , \\
I_{d-1} +  \l_{d-1} - \, d \k_{d-2} 
\end{pmatrix}\  .
\ea
As before, if a nonvanishing action is to exist, one must have $(J_d, I_{d-1})$ non-trivial under this equivalence.

Now consider the effect of a one-parameter family of superdiffeomorphisms $f_t: \cM \ra \cM$ as before. As we saw in Section \ref{Superfields with boundaries}, we cannot expect that the action on a manifold with boundary will be invariant under the full set of superdiffeomorphisms. Instead, it can only be invariant  under a subset of diffeomorphisms such that both single transformations and composites of transformations preserve the bosonic normal to the boundary. This condition restricts us to diffeomorphisms generated by a subspace of supervectors $\cV \subset T \cM$ where
\ba
i_{v^\tb{1}} n &= 0\ ,  & i_{v^\tb{2}} n & = 0\ ,  &  i_{[v^\tb{1}, v^\tb{2} \} } n & = 0\ ,   & \forall \quad v^\tb{1}, v^\tb{2} &\in \cV \  , 
\label{vFundConstraints}
\ea
where $n = E^a n_a $ is the outward-pointing bosonic unit normal form. These constraints imply that 
\ba
 v^{\tb{1} A}  v^{\tb{2} B} T_{A B}{}^C n_C = 0\  . 
 \label{vQuadConstraint}
 \ea
which we will refer to as the quadratic constraint.

Under the surviving diffeomorphisms, the transformed action is given by 
\ba
S_t = \int_{\cM_0} s^* \circ f^*_t J_d - \int_{\pa \cM_0} c^* \circ s^* \circ f^*_t I_{d-1}\ . 
\ea
Thus, the action will be supersymmetric if 
\ba
0 & = \fr{d S_t}{d t }   = \int_{\cM_0} s^* \cL_v J_d -  \int_{\pa \cM_0} c^* \circ s^* \cL_v I_{d-1}\ ,  \nn \\
& = \int_{\cM_0} s^* ( d i_v J_d + i_v d J_d ) -  \int_{\pa \cM_0} c^* \circ s^*( d i_v I_{d-1}  + i_v d I_{d -1} )\  ,  \nn \\
& = \int_{\cM_0} (d (s^* i_v J_d ) + s^* i_v d J_d )-  \int_{\pa \cM_0} ( d ( c^* \circ s^* i_v I_{d-1}  )+ c^* \circ s^* i_v d I_{d -1} )\  ,  \nn \\
& = \int_{\cM_0} s^* i_v d J_d + \int_{\pa \cM_0} c^* \circ s^* i_v ( J_d  -   d I_{d -1} )\  . 
\ea
To solve this, both the first and the second term must vanish separately. The vanishing of the first term is achieved by imposing $dJ_d = 0 $ as before. This means that when boundary effects are included, the bulk action is constructed exactly in the same way as when they are ignored. Clearly, the vanishing of the second term places some constraints on $I_{d-1}$. On the surface, it would appear that one requires $J_d = d I_{d-1}$. However, this constraint forces $J_d$ and $I_{d-1}$ to be exact under the equivalence \eqref{relcohomology1} and so makes the action vanish. In fact this constraint is not required as $v$ is not a general vector but one which must satisfy \eqref{vFundConstraints}. To proceed, we now impose $dJ_d = 0$ and consider the second term in more detail:
\ba
\fr{d S}{dt} = &\int_{\pa \cM_0} c^* \circ s^* i_v (J_d - dI_{d-1}) = 0\ . 
\ea
In the rigid case, this implies
\ba
 n \wedge \Big[i_v(J - dI) \Big]_{d-1,0} = 0\ . 
\ea
Then, expanding out $J_d-dI_{d-1}$ in terms of its super $(p,q)$-form parts, we find 
\ba
 (i_{v} n ) (J - dI)_{d,0} + n \wedge \big ( i_{v_{0,1}} (J -  dI)_{d-1,1} \big)  = 0\ . 
\ea
The first term of this vanishes upon using \eqref{vFundConstraints} and the remaining constraint is expressed in components as 
\ba
n_{a_1} v^{\a} \e^{a_1 \dots a_d} \big (J_{\a a_2 \dots a_d } -  d \na_{ [\a}  I_{ a_2 \dots a_d )} - \fr{d(d-1)}2 T_{[ \a a_2| }{}^B I_{B | a_3 \ldots  a_d )}  \big ) = 0\  .
\label{dScondition}
\ea
The discussion in the local case is analogous, in which case \eqref{dScondition} becomes
\ba
n_{m_1} v^{\a} \e^{m_1 \dots m_d} E_{m_2}{}^{A_2}  \ldots E_{m_d}{}^{A_d} \big (J_{\a A_2 \dots A_d } -  d \na_{ [\a}  I_{ A_2 \dots A_d )} - \fr{d(d-1)}2 T_{[ \a A_2| }{}^B I_{B | A_3 \ldots  A_d )}  \big ) = 0\  .
\label{dSconditionlocal}
\ea
By making an ansatz for $I_{d-1}$ which solves these equations, we can find the appropriate boundary completion to a given bulk action, which we will demonstrate with some examples in Section \ref{examples}. 

\subsection{Relative Cohomology and the Superboundary}
\label{superboundaries}

We now consider an alternative perspective on the above construction involving relative cohomology, which is helpful in illuminating additional structure in the bulk + boundary system. To do this, we recall\footnote{For an application of relative cohomology to the problem of large gauge transformations in 2-brane and 5-brane M-theory backgrounds, see Ref.\ \cite{Kalkkinen:2002tk}.} that, given a $d$ dimensional manifold $\cA$ and an inclusion $\i: \cB \rightarrow \cA$ of a $d-1$ dimensional submanifold $\cB \subset \cA$, one can define
\ba
	\O^*(\i) &= \sum_p \O^p (\i)\ ,  & &  \text{where} &\O^p (\i) &= \O^p (\cA) \oplus \O^{p-1}(\cB)\ , 
\ea
and where $\O^p(\cA)$ is the set of p-forms on the manifold $\cA$. Then, considering $( A_p,  B_{p-1} ) \in \O^p(\i) $ one can define a natural exterior derivative $\dc : \O^p(\i) \ra \O^{p+1}(\i)$ given by
\ba
	\dc (A_p , B_{p-1} ) = ( d A_p, \i^* A_p - d B_{p-1})\  . 
\ea
The relative cohomology group $H^p (\cA , \cB)$ is then defined to be the usual quotient of the $\dc$-closed forms in $\O^p(\i)$ by the $\dc$-exact ones. A $\dc$-closed element $( A_p , B_{p-1})$ has
\ba
	 d A_p &= 0\ ,  &  \i^*  A_p & = d B_{p-1}\ , 
\label{relClosed}
\ea
so $A_p$ is closed and the pullback of $A_p$ onto $\cB$ is exact, while a $\dc$-exact element has
\ba
	 A_p & = d C_{p-1}\ ,  &   B_{p-1} & =  \i^* C_{p-1} - d D_{p-2}\ . 
\ea

Next we can define the integral of $(A_d, B_{d-1})$ over the pair of spaces $(\cA, \cB)$ by 
\ba
\int_{(\cA, \cB)} (A_d, B_{d-1})  = \int_\cA A_d - \int_\cB B_{d-1}\  . 
\ea 
Next we define the relative boundary operator $\pac$ to act on the pair $(\cA, \cB)$ as
\ba
\pac (\cA, \cB) = (\hat \i (\cB) - \pa \cA, \pa \cB)\ , 
\ea
where $\hat \i : \cB \ra \pa \cA$ such that $c_{\pa \cA} \circ \hat \i = \i $ and $c_{\pa \cA} : \pa \cA \ra \cA $. Using the standard version of Stokes' theorem, 
\ba
\int_\cA d A_{d-1} = \int_{\pa \cA} c_{\pa \cA}^* A_{d-1}\ , 
\ea
one can show that the generalised Stokes' theorem relating $\pac$ and $\dc$ is
\ba
\int_{(\cA, \cB)} \dc (A_{d-1}, B_{d-2})  = - \int_{ \pac ( \cA, \cB )} ( c^*_{\pa \cA} A_{d-1}, c^*_{\pa \cB}B_{d-2} )\  .
\label{relStoaks}
\ea

If we now consider the case where $\pa \cA = \cB$ and let $c = c_{\pa \cA} = \i$, $\hat \i = \id$, $c_{\pa \cB} = 0$ we can construct an integral of the form 
\ba
S[ (A_d, B_{d-1} ) ]  = \int_{(\cA, \pa \cA)}  (A_d, B_{d-1})  \ .\label{genrelint}
\ea
Then, if $(A_d,  B_{d-1}) $ is $\dc$-exact, one finds
\ba
	S[ \dc (C_{d-1}, D_{d-2} ) ] = \int_{\pac ( \cA, \pa \cA ) } ( c^* C_{d-1} , 0 )  = 0\ . 
\ea
Consequently, the integral \eqref{genrelint} depends only on the relative cohomology class of $ (A_d , B_{d-1}) $:
\ba
	S[( A_d, B_{d-1}) ] = S[ ( A_d + d C_{d-1} , B_{d-1} + c^* C_{d-1} - d D_{d-2} )]\ . 
\ea
If  $\cA = \cM_0 $,  $\cB = \pa \cM_0$, $A_d = s^* J_d$ and $B_{d-1} =  c^* \circ s^* I_{d-1}$, then this becomes equivalent to \eqref{relcohomology1}. 

Next we define the ``superboundary''\footnote{The notion of a boundary superspace has appeared previously in the context of two-dimensional supersymmetry in \cite{Nevens:2006ht}.} $\til\cM$ to be the manifold with $d -1$ bosonic dimensions and $\fr{n}{2}$ fermionic dimensions given by the locus of the boundary $\pa \cM_0$ under the surviving supersymmetry transformations \eqref{vFundConstraints}.
Then, choosing $z^M$ to be coordinates on $\cM$ and $z^{\til M}$ to be coordinates on $\til \cM$, we may define the embedding matrix
\ba
E_{\til A}{}^A = E_{\til A}{}^{\til M} \pa_{\til M} z^M E_{M}{}^A . 
\ea
We can then follow the standard description of one supermanifold embedded into another one \cite{Howe:1996mx}. We thus impose the condition that the odd tangent space of $\til \cM$ lies within the odd tangent space of $\cM$.  This implies that the embedding matrix satisfies
\ba
E_{\til \a}{}^a = 0\  . 
\ea
Combining this with the defining equation for the torsion and extracting the dimension-zero part, we obtain 
\ba
E_{\til \a}{}^\a E_{\til \b}{}^\b  T_{\a \b}{}^c = T_{\til \a \til \b}{}^{\til c} E_{\til c}{}^c\ , 
\ea
where $T_{\til \a \til \b}{}^{\til c}$ is the dimension-zero part of the torsion on $\til \cM$. Contracting the indices on this equation with two fermionic vectors $\til v_1^{\til \a}$ and  $\til v_1^{\til \a}$ defined on $\til \cM$ and with the bosonic normal $n_a$ gives 
\ba
\til v_1^{\til \a} \til v_2^{\til \a} E_{\til \a}{}^\a E_{\til \b}{}^\b  T_{\a \b}{}^c n_c = 0 \ ,
\ea
where we have used $E_{\til c}{}^c n_c = 0$. This shows that the quadratic constraint \eqref{vQuadConstraint} is satisfied if
\ba
v^\a = \til v^{\til \a} E_{\til \a}{}^\a
\label{vSolTil}
\ea
where $\til v^\til \a$ is any odd supervector on $\til \cM$. 

Next let $\til s : \pa \cM_0 \ra \til \cM$ and $\til c: \til \cM \ra \cM$ in such a way that we have a commuting diagram of maps: 
\ba
\begin{xy} (0,20)*+{\pa \cM_0}="a"; (20,20)*+{\cM_0}="b"; 
	(0,0)*+{\til {\cM} }="c"; (20,0)*+{\cM}="d";
	{\ar^c "a";"b"};
	{\ar_{\til{s}} "a";"c"};
	{\ar^{s} "b";"d"};
	{\ar^{\til{c}} "c";"d"};
\end{xy}
\ea
with corresponding pullbacks that satisfy $\til s^* \circ \til c^* = c^* \circ s^*$. As before, we define a bulk superform $J_d$ on $\cM$ but we now define the boundary superform $\til I_{d-1}$ on $\til \cM$.  The action we consider is then given by 
\ba
S = \int_{(\cM_0, \pa \cM_0)} (s^* J_d, \til s^* \til I_{d-1} )\ . 
\label{defStil}
\ea
Under an odd superdiffeomorphism, this is transformed to
\ba
S_t = \int_{(\cM_0, \pa \cM_0)} (s^* \circ f^*_t J_d, \til s^* \circ \til f^*_t \til I_{d-1} )
\  ,
\label{varStil}
\ea
where $f^*_t : \cM \ra \cM$ is a one-parameter family of superdiffeomorphisms generated by $v^\a$ satisfying \eqref{vSolTil} and $\til f^*_t : \til \cM \ra \til \cM$ is a one-parameter family of superdiffeomorphisms generated by $\til v^{\til \a} $. Proceeding as before, one can show that the action is supersymmetric if
\ba
0 = \fr{d S_t}{d t }  = \int_{( \cM_0, \pa \cM_0 ) } \big ( s^* i_v d J_d , \, \til s^*  i_{\til v} (   d \til I_{d -1} - \til c^*   J_d  ) \big) \ ,
\ea
which implies that 
\ba
dJ_d  & = 0\ , &  \til c^* J_d &=  d \til I_{d-1} &&\text{so} &d^{( \til c )} ( J_d , \til I_{d-1} ) &= 0 \ .
\label{JIrel}
\ea
 We consequently find that the pair $( J_d, \til I_{d-1} )$ is an element of the relative cohomology group $H^d( \cM, \til \cM)$ and that the invariant action is the natural generalisation in the relative cohomology framework of the case without a boundary. 

\section{ Examples} 
\label{examples}

\subsection{ 3D and 4D Ectoplasm with an Edge} 

As an example of the construction given in Section \ref{Superforms with boundaries}\,, we now consider the rigid 3D case, for which the only nonvanishing component of the torsion is $T_{\a \b}{}^c = i  \g^c{}_{\a \b}$. Accordingly, we solve the constraints \eqref{vFundConstraints} by imposing 
\ba
n_a v^a & = 0\ , & n_a \g^a{}_\a{}^\b v_\b = v_\a\ . 
\label{3DvConstraint}
\ea
 We know that $J_3$ is given by \eqref{3DJ} and so we make the ansatz that $I_2$ is given by 
\ba
I_{a b} & =  -   \e_{a b c} n^c J_0\ ,  & I_{\a b} & = 0\  . 
\label{3DK}
\ea
Substituting this into \eqref{dScondition}, we find that
\ba
0 & =\e^{a b c} n_c v^\b ( \e_{a b d} \g^{d}{}_{\b}{}^\a D_{\a} J_0  +  \e_{ a b d} n^d D_\b J_0 )\  ,
\ea
which can be rearranged to give
\ba
0 & = D^\a J_0 ( v_\a - \g^a{}_\a{}^\b n_a v_\b )\  . 
\ea
This is then satisfied by imposing the constraint \eqref{3DvConstraint}. Substituting the derived values of $J_d$ \eqref{3DJ} and $I_{d-1}$ \eqref{3DK} into \eqref{defI}, we find that the boundary modified F-density is given by 
\ba 
\int_{\cM_0}  d^3 x\, i  D^\a D_\a J_0 \big |  + \int_{\pa \cM_0} d^2 x J_0\big |\  . 
\ea
 
As another example, consider the rigid 4D case. Here, the nonvanishing components of the torsion are are ($T_{\a \dot \b}{}^c = i \s^c{}_{\a \dot \b}$\,, $T_{\dot \a \b}{}^{c} =  i \bar \s^{c}{}_{\dot \a \b}$), so we solve \eqref{vFundConstraints} by imposing
\ba
n_a v^a & = 0\ , & \s^a{}_{ \a}{}^{\dot \a}  n_a \bar v_{\dot \a}  & = v_\a\ . 
\label{4DvConstraint}
\ea
Then, by considering the form of \eqref{dScondition} and the superform \eqref{4DJ}, we make the following ansatz for $I_3$:
\ba
I_{a b c} & = -   \e_{a b c d} n^d ( J_0 + \bar J_0 )\ ,   & I_{\a b c} & = 0\ . 
\label{4DI}
\ea
Substituting this into  \eqref{dScondition}, we find 
\ba
0  = \ & \e^{a b c d} n_d v^\a ( \e_{a b c e} \s^e{}_{\a \dot \a} \bar D^{\dot \a} \bar J_0 -  \e_{a b c e } n^e D_\a J_0) \nn \\& 
+   \e^{a b c d} n_d \bar v^{\dot \a} (  \e_{a b c e} \bar \s^e{}_{\dot \a \a} D^{ \a} J_0 -  \e_{a b c e } n^e \bar D_{\dot \a}  J_0)\  ,
\label{4DdI1}
\ea
which can be rearranged to give 
\ba
0 =  D^{ \a} J_0  ( \s^e{}_{ \a}{}^{\dot \a}  n_e \bar v_{\dot \a}  - v_{\a} ) 
+ \bar D^{\dot \a}  \bar J_0  ( \bar \s^e{}_{\dot \a}{}^\a n_e v_\a - \bar v_{\dot \a} )\ .
\ea
As before, this is satisfied by imposing the constraint \eqref{4DvConstraint}. Rewriting \eqref{defI}  using \eqref{4DJ}  and \eqref{4DI}, we find
\ba
S &= \int_{\cM_0} d^4 x \fr12 i  \bls D^\a D_\a J_0 \big | + \bar D^{\dot \a} \bar D_{\dot \a} \bar J_0 \big |\brs   +  \int_{\pa \cM_0} d^3 x \bls J_0 \big | +  \bar J_0 \big | \brs\  . 
\ea
This reproduces the rigidly supersymmetric results of references \cite{Belyaev:2007bg} and \cite{Belyaev:2008xk}. In a similar fashion, one can use the present method to deduce the appropriate boundary modification to any superform action and hence obtain the corresponding  ``SUSY without b.c.'' superfield action.  

\subsection{ 5D Ectoplasm with an Edge} 

Having reproduced the known rigid ``SUSY without b.c.'' results, we can use the present ``Ectoplasm with an Edge'' formalism to derive the boundary modifications to the rigid F-density in 5D supersymmetry. As far as we know, this will be a new result. In standard 5D superspace without boundaries, actions can be built by considering a linear superfield $J_{i j}$, which satisfies by definition
\ba
J^{i j} &= J^{(i j)}\ ,  &  D_\a^{(i} J^{j k) } & = 0\ , 
\ea
where $i$ is an $SU(2)$ doublet index with respect to which the spinors are symplectic-Majorana:
\ba
\bar {v_{\a}^i}  = - C^{\a \b} \e_{ i j} v_\b^j\ . 
\label{SMCondition} 
\ea
We take the only non-vanishing component of the torsion to be $T_{\a i \b j}{}^c =i  \g^c{}_{\a \b} \e_{i j}$. The standard F-density then reads 
\ba
S &= \int_{\cM_0} d^5 x i D^{ \a i } D_\a^{\ j} J_{ i j} \big |\ . 
\label{5DAct}
\ea
To find the appropriate boundary modification to this action, we begin by finding the relevant closed super 5-form. The construction of this superform is different from the cases considered previously as it turns out that the closed super 5-form that we will construct is of a particular type known as a Chern-Simons superform. 

We begin the construction by considering the closed super 4-form given by
\ba
X_{\a i \b j \g k \l l} & = 0\ , & 
X_{a \a i \b j \g k } & = 0\ , \nn \\
X_{a b \a i \b j } &= -12 i \g_{a b \, \a \b} J_{i j}\ ,  &
X_{a b c \a i} &=  - 4   \g_{a b c}{}_\a{}^\b D_\b^j  J_{i j}\ , \nn \\
X_{a b c d} &=  - i  \g_{a b c d}{}^{\a \b} D_{\a}^i D_\b^j J_{i j } \ .
\ea
Using Poincar\'e's lemma, we can then write $X_4 = d Q_3$, where generally $Q_3$  cannot be expressed solely in terms of $J_{ij}$ and its derivatives. For example, in the case where the action describes the kinetic terms of 5D super Yang-Mills,  $X_4 = \tr F_2 F_2$, where $F_{2,0} |$ is the Yang-Mills field strength, so $J_{ij}$ is gauge invariant. However, $Q_{3,0} |$ is the Chern-Simons 3-form, which is not gauge-invariant and cannot be built solely from $J_{i j}$. 

We next proceed by considering the super 6-form described by $W_{6} = C_{0,2} X_4$, where $C_{0,2} = \fr12 E^{\b j} E^{\a i} C_{\a \b} \e_{i j} $. This can clearly be written as $W_6 = d Z_5 $, where $Z_5 = C_{0,2} Q_3$. However, it can also be written in the exact form $W_6 = d K_5$, where $K_5$ is gauge invariant:
\ba
K_{\a i \b j \g k \l l \d m} & = 0\  , & 
K_{a \a i \b j \g k \l l } & = 0\ ,  \nn\\
K_{a b \a i \b j \g k } & = 0\ , & 
K_{a b c \a i \b j } & = - 12  \g_{a b c \, \a \b} J_{i j}\ ,  \nn \\
K_{a b c d \a i } & =  4 i\g_{a b c d}{}_{\a}{}^\b D_{\b}^j J_{i j}\ ,  & 
K_{a b c d e} & =  - i \e_{a b c d e } D^{\a i}  D_{\a}{}^{ j} J_{ij}\ . 
\label{5DJ}
\ea
The possibility of writing $W_6$ both as $dZ_5$ and as $dK_5$ is known as Weil triviality \cite{Bonora:1986xd}.
In consequence of the Weil triviality, we can form a closed super 5-form from the difference: $J_5 = K_5 - Z_5$, the integral of which gives the action \eqref{5DAct}\footnote{An alternative for $Z_5$ can be constructed in a chosen $\q_{\a i}$ frame as $ Z_5 = \fr12 \q_{0,1} X_4 $, where $\q_{0,1} = E^{\a i} \q_{\a i} $. This can easily be shown to satisfy $d Z_5 = W_6$, but it does not transform as a superform under superdiffeomorphisms owing to the explicit $\q_{0,1}$ term. }.

As before, the presence of a boundary in the spacetime manifold partially breaks the supersymmetry, forcing us to impose conditions \eqref{vFundConstraints}. One might think that, by analogy with the 3D and 4D examples, the quadratic constraint \eqref{vQuadConstraint} could be solved by imposing 
\ba
n_a \g^{a}{}_{\a}{}^\b v_{\b i} & = v_{\a i}\ . 
\label{5DvConSol1} 
\ea
However, this is not the case since imposing \eqref{SMCondition} and \eqref{5DvConSol1}  simultaneously would imply that the spinor $v_\a^i$ is identically zero. Instead, we solve \eqref{vFundConstraints} by imposing
\ba
n_a v^a & = 0\ , &  m_I n_a \s^I{}_i{}^j \g^{a}{}_{\a}{}^\b v_{\b j} & = v_{\a i}\ , 
\label{5DvConSol2} 
\ea
where $I = 1,2,3$ is an $SU(2)$ triplet index and $\s^I{}_i{}^j$ are the Pauli matrices; $m_i$ is an arbitrary $SU(2)$ normalised constant triplet. Combining this with \eqref{SMCondition} now does not imply that $v_\a^i = 0$. We see from this that the introduction of the boundary has not only broken 5D supersymmetry and Poincar\'e symmetry, but it has also broken the $SU(2)$ symmetry by forcing us to introduce the vector $m_I$ which picks out a fixed direction in $SU(2)$. 

We thus propose that the boundary superform is given by 
\ba
I_{a b c d} & =  6  \e_{a b c d e} n^e m_I \s^{I i j} J_{i j}\  , & I_{\a i b c d} & = 0\  ; 
\label{5DK} 
\ea
we then find that substituting \eqref{5DJ} and \eqref{5DK} into the requirement \eqref{dScondition} gives
\ba
0 = \e^{a b c d e} n_e v^{\a i } ( 4 i \g_{a b c d}{}_\a{}^\b D_\b^j J_{ij}  -6 \e_{ a b c d f} n^f m_I \s^{I j k} D_{\a i} J_{j k}  ) \  , 
\ea
which can be rearranged to give 
\ba
0 = D^{\a j } J_{i j} ( n_a \g^a{}_\a{}^\b v_\b^i + m_I \s^{I i}{}_k v_\a^k )\ .
\ea
This is indeed satisfied by imposing \eqref{5DvConSol2}. We therefore find that the 5D rigid boundary modified F-density is given by
\ba
S &= \int_{\cM_0} d^5 x i  D^{ \a i } D_{\a}^{ j} J_{ i j} \big | + \int_{\pa \cM_0} d^4 x 6 m_I \s^{I i j} J_{i j}  \big |\  . 
\ea

\section{Conclusion}

In this paper, we have shown how the ectoplasm formalism for the construction of supersymmetric invariants via closed forms in superspace can be adapted to supersymmetric systems on manifolds with boundaries. A natural extension of this to local supersymmetry will be considered in a separate paper. 

An important question that we have not addressed here is whether such a formalism can be applied to theories for which no off-shell formalism is available, such as maximal super Yang-Mills and maximal supergravity. The ectoplasm formalism has proven useful in characterising the properties of candidate ultraviolet counterterms in these theories \cite{Bossard:2009mn,Bossard:2010bd,Bossard:2010pk}, despite the on-shell character of their supersymmetries. An intermediate situation exists for systems formulated in harmonic superspace, involving an infinite number of auxiliary fields,  such as 8-supercharge hypermultiplet models and fully supercovariantised 8-supercharge SYM.  These have appropriate harmonic superspace formulations in four \cite{Galperin:1984av} and six dimensions \cite{Howe:1985ar}, and have been discussed ectoplasmically in \cite{Biswas:2001wu}. The great advantage of all off-shell formalisms for theories on manifolds with boundaries is that one can straightforwardly include extra boundary matter supermultiplets. Their boundary actions in an off-shell formulation are separately supersymmetric, although they may also need to be covariantised by couplings to bulk gauge and supergravity fields.

An open question concerns the adaptation of the present boundary formalism to situations without a full off-shell supersymmetry formulation. Given the reduction in unbroken supersymmetry on the boundary, one possibility might be to generalise the present construction to cases where just the surviving supersymmetry is realised off-shell.

In this paper, we have also developed the relation between the boundary theory formulation and the superembedding formalism which has been employed in the study of worldvolume theories of supersymmetric $p$-branes \cite{Bandos:1995zw,Howe:1996mx,Howe:1997wf,Sorokin:1999jx,Bandos:1995dw,Howe:1998tsa}. In the superembedding program, one implements the requirements of $\kappa$-symmetry in a natural geometrical context by the embedding conditions, commensurate with a reduction in the degree of unbroken supersymmetry on the brane worldvolume. In a boundary theory construction, this corresponds to the reduction in unbroken supersymmetry on the boundary. The superembedding formalism describes the 
natural dynamics of a brane itself, \ie the dynamics of the Goldstone multiplet corresponding to its broken translation symmetry, supersymmetry and their superpartners. In the present paper, we have not focused attention on this Goldstone supermultiplet. But in braneworld and boundary contexts, such Goldstone supermultiplets play key r\^oles \cite{Lukas:1998yy,Lukas:1998tt,Lehners:2005su,Lehners:2002tw}, and it will be important to distinguish the dynamics of these ``Goldstone'' multiplets from the other dynamics on the brane or boundary surfaces.

\section*{Acknowledgments}

KSS would like to thank the Kavli Institute for Theoretical Physics for hospitality during the course of this work, and for support in part by the National Science Foundation under Grant No.\  NSF PHY05-51164. The work of KSS was also supported in part by by the STFC under rolling grant PP/D0744X/1.

\begin{appendix}

\section{Conventions}\label{conventions}

Throughout this paper, we use the convention that $M = 1, \ldots d+n$ is a supermanifold world index with $A = 1, \ldots d +n $ an index on the supermanifold tangent space. Similarly $m = 0, \ldots d-1 $ is  a bosonic submanifold world index with $a = 0, \ldots d -1 $ an index on the bosonic submanifold tangent space. In our general discussions here, we will take $\m = 1, \ldots n$ to be a fermionic world index and $\a = 1 \ldots n$ to be a fermionic tangent space index. We make use of a superspace covariant derivative defined by 
\ba
\na_A E_{M}{}^B & = \pa_A E_M{}^B + \O_{AC}{}^B E_{M}{}^C, & \na_A E_B{}^M & = \pa_A E_B{}^M  -  \O_{AB}{}^C E_C{}^M.& 
\ea 
Then the torsion is defined in terms of this connection as 
\ba
 T_{A B}{}^C =  - 2 \na_{[A} E_{B)}{}^M E_M{}^C = 2 \O_{[A B)}{}^C - 2  \pa_{[A|} E_{|B)}{}^M  E_{M}{}^C . 
\label{defTorsion}
 \ea
where $\pa_A = E_A{}^M \pa_M$ and $E_{M}{}^A$ is the supervielbein . In the rigid cases that we consider, the only nonvanishing component of the torsion is $T_{\a \b}{}^c = i \g^c{}_{\a\b}$  and the connection vanishes, $\O_{AB}{}^C = 0$. Then $\na_A$ is simply given by 
\ba
\na_a &= E_a{}^M \pa_M = \pa_a &  \na_\a & = E_\a{}^M \pa_M = \pa_\a -  \fr12 i  \g^c{}_{\a \b} \q^\b \pa_c = D_\a\ ,
\ea
where $D_\a$ is the standard flat-superspace covariant derivative. Here, we have used conventions in which $\q^\a$ is real, which means that since $(\q^\a \q^\b)^* = \q^\b \q^\a$ it follows that $\pa_\a$ is imaginary. 

In rigid 3D supersymmetry, we use the conventions
\ba
\e_{\a \b} & = \e^{\a \b}, & \e^{12} & = 1\ ,
\ea
\ba
\e_{\a \b} \e^{\b \g}&= - \d_\a{}^\g\ , & \p^\a & = \e^{\a \b} \p_\b\ , &  \p_\b & = \p^\a \e_{\a \b}\ ,
\ea
\ba
E^A J_A & = E^a J_a + E^\a J_\a\ , &   [\q^ \a]^* & = \q^{ \a}, & [D_ \a]^* & = - D_{\a}\ , 
\ea
\ba
\{ \g_{a},  \g_b \}_\a{}^\b & = \d_\a{}^\b \h_{a b}\ , & \h_{a b } & = \text{diag} ( -1, +1, +1 )\ , & \g^{a b c}{}_\a{}^\b &= \e^{a b c} \d_\a{}^\b\ ,
\ea
\ba
 \e_{a b c} \e^{d  e f} & = - 3! \d_{[a}{}^d \d_{b}{}^e \d_{c]}{}^f, & \e^{0 1 2} & = 1\ ,
\ea
\ba
\{ D_{\a},  D_{\b}  \} & = -  i \g^c{}_{ \a \b} \pa_c\ , & T_{\a \b}{}^c = i  \g^c{}_{\a \b}\ .
\ea
This implies the useful identities
\ba
 \g_a{}_{(\a \b} \d_{\g)}{}^\d &= \e_{abc} \g^b{}_{(\a \b} \g^c{}_{\g)}{}^\d\ ,&
D^\b D_\a D_\b = \{ D_\a , D^\b D_\b \} &= 0\ .
\ea

In rigid 4D supersymmetry, we use the conventions 
\ba
\e_{\a \b} \e^{\b \g} &= - \d_{ \a}{}^{ \g}\ , & \p^{\a} & = \e^{\a \b} \p_\b\ , &  \p_\b & = \p^\a \e_{\a \b}\ ,  
\ea
\ba
\e_{\dot \a \dot \b} \e^{\dot \a \dot \g} &= - \d_{\dot \a}{}^{\dot \g}\ , &\bar  \p^{\dot \a} & = \e^{\dot \a \dot \b} \bar \p_{\dot \b}\ , &  \bar \p_{\dot \b} & = \bar \p^{\dot \a} \e_{\dot \a \dot \b}\ , 
\ea
\ba 
[\e_{\a\b}]^* &= \e_{\dot\a \dot\b}\ , &[\e^{\a\b}]^* =\e^{\dot\a \dot\b}\ ,
\ea
\ba
\bar\s^a{}^{\dot\a \a} &= \e^{\a \b} \e^{\dot\a \dot\b} \s^a{}_{\b \dot\b}\  , &
\left[ \s^a{}_{\a \dot\a} \right]^* &= \bar\s^a{}_{\dot\a \a}\ , 
\ea
\ba
E^A J_A & = E^a J_a + E^\a J_\a + E^{\dot \a} J_{\dot \a}\ , & [\q^ \a]^* & = \bar \q^{\dot \a}\ , & [D_ \a]^* & = - \bar D_{\dot \a}\ , 
\ea
\ba
\s^{a}{}_{ \a}{}^{ \dot \b} \bar \s^b{}_{\dot \b \b}  & = \h^{a b} \e_{\a \b} + \s^{a b}{}_{ \a \b}\ , & \h_{a b} & = \text{diag} ( -1, +1, +1, +1 )\  , 
\ea
\ba
\s^{ab}{}_{ \a \b} &= -\frac{i}{2} \e^{abcd} \s_{cd \a \b }\ , & \bar\s^{ab}{}_{ \dot \a \dot \b} &= +\tfrac{i}{2} \e^{abcd} \bar\s_{cd \dot \a \dot \b }\  , 
\ea
\ba
\e_{a b c d} \e^{e f g h} & = - 4! \d_{[a}{}^e \d_{b}{}^f \d_{c}{}^{g} \d_{b]}{}^h ,  & \e^{0123} = + 1\ , 
\ea
\ba
\{ D_{\dot \a},  D_{\b} \}& = -  i \bar \s^c{}_{ \dot \a \b} \pa_c\ , & T_{\dot \a \b}{}^c &= i \bar \s_{\dot \a \b}^c\ ,  & T_{\a \dot \b}{}^c & = i \s^c{}_{\a \dot \b} \ .
\ea
which implies the useful identities
\ba
\bar\s^{[a}{}_{\dot\g \a} \s^{b]}{}_{\b \dot\d} 
	&= - \frac{1}{2} \s^{ab}{}_{\a \b} \e_{\dot\g \dot\d} + \tfrac{1}{2} \bar\s^{ab}{}_{\dot\g \dot\d} \e_{\a \b}\  , & 
D_\a D_\b & = -\frac{1}{2} \e_{\a \b} D^\g D_\g\ ,  & D_\a D_\b D_\g &= 0 \ .
\ea

In rigid 5D supersymmetry, we use the conventions
\ba
\e_{i j} & = \e^{i j }\ , & \e^{12} & = 1\ , & C^{\a \b} = - C^{\b \a}\  ,
\ea
\ba
\e_{i j } \e^{j k} &= - \d_i{}^k\ , & \p^i_\a & = \e^{i j} \p_{ \a j }\  , &  \p_{\a j} & = \p^i_\a \e_{i j}\ ,
\ea
\ba
C_{\a \b} C^{\b \g} &= - \d_\a{}^\g\ , & \p^\a_i  & = C^{\a \b} \p_{\b i}\ , &  \p_{\b i } & = \p^\a_i  C_{\a \b}\ , 
\ea
\ba
E^A J_A & = E^a J_a + E^\a J_\a\ , &   [\q^ \a]^* & = \q^{ \a}\ , & [D_ \a]^* & = -  D_{\a}\  , 
\ea
\ba
\{ \g_{a}\ ,  \g_b \}_\a{}^\b & = \d_\a{}^\b \h_{a b}\ ,& \h_{a b } & = \text{diag} ( -1, +1, +1, +1, +1 )\ ,& \g^{a b c d e}{}_\a{}^\b &= i \e^{a b c d e} \d_\a{}^\b\ ,
\ea
\ba
\e_{a b c d e } \e^{f g h i j} & = - 5! \d_{[a}{}^f \d_{b}{}^g \d_{c}{}^h\d_{d}{}^i \d_{e]}{}^j\ ,& \e^{0 1 2 3 4} & = 1\ ,
\ea
\ba
\{ D_{\a i },  D_{\b j}  \} & = - i    \g^c{}_{ \a \b} \e_{i j}  \pa_c\ ,& T_{\a \b}{}^c =i  \g^c{}_{\a \b} \e_{i j} \ .
\ea
which implies the useful identities
\ba
\g_a{}_{\a \b} \g^{ab}{}_{\g \d} &= 2 \Big[ \g^b{}_{\g [\a} C_{\b] \d} + ( \g \lra \d ) \Big]\  ,\nn \\
\g_c{}_{\a \b} \g^{cab}{}_{\g \d} &= -\frac{3}{4} \Big[ \g^{ab}{}_{(\b \g} C_{\d) \a} - (\a \lra \b) \Big] -\frac{9}{4} \Big[ \g^{ab}{}_{\d [\a} C_{\b \g]} + (\g \lra \d) \Big] \ ,\nn \\
\g^{[a}{}_{\a \b} \g^{b]}{}_{\g \d} &= - \frac{3}{2} \Big[ \g^{ab} \, {}_{\d [\a} \, C_{\b \g]} -(\g \lra \d ) \Big]\  ,
\ea
and also implies that for a Linear superfield satisfying $D_{\a ( i} J_{j k) } = 0 $
\ba
D_\a^k D_{\b k} J_{i j}  = i \g^a{}_{\a \b} \pa_a J_{i j} \ . 
\ea

\section{ $J_d$ closure conditions}\label{closure conditions}

The condition that the 3D super 3-form be closed implies the non-trivially satisfied constraints
\ba
T_{( \a \b}{}^c J_{\g \d ) c} &= 0 
\ea
\ba
\na_{[a} J_{b] \a \b }  + \na_{(\a} J_{\b)a b} = - \fr12 T_{\a \b}{}^c J_{a b c} 
\ea
\ba
3 \na_{[a} J_{ b c ] \a}  - \na_\a J_{a b c} = 0 \ .
\ea

The condition that the 4D super 4-form be closed implies the non-trivial constraints
\ba
T_{\dot \a ( \b}{}^c J_{\g \d ) c d } & = 0 
\ea
\ba
\bar \na_{\dot \a} J_{\a \b c d } = 2 T_{\dot \a ( \a}{}^e J_{\b ) c d e} 
\ea
\ba
2 \na_{(\a} J_{\b) c d e} + 3 \na_{[c} J_{d  e] \a \b} = 0 
\ea 
\ba
\bar \na_{\dot \a} J_{\a b c d} + \na_\a J_{\dot \a b c d} & = T_{\dot \a \a}{}^e J_{b c d e} 
\ea
\ba
\na_\a J_{b c d e} + 4 \na_{[b} J_{c d e] \a} = 0 \ .
\ea

The condition that the 5D super 4-form $X_4$ be closed implies the non-trivial constraints 
\ba
T_{ ( \wh \a \wh \b }{}^b X_{\wh \g \wh \l ) a b } & = 0 
\ea
\ba
\na_{(\wh \a} X_{\wh \b \wh \g) d e} =  T_{( \wh \a \wh \b }{}^f X_{\wh \g) d e f }
\ea
\ba
2 \na_{(\wh \a} X_{\wh \b ) c d e } + 3 \na_{[ c} X_{d e ] \wh \a \wh \b }=  T_{\wh \a \wh \b }{}^f X_{ c d e f }
\ea
\ba
\na_{\wh \a}  X_{ b c d e }  + 4 \na_{[b} X_{c d e ] \wh \a }  = 0
\ea
\ba
\na_{[a } X_{b c d e ]}= 0  
\ea
where $\wh\a$ is a bi-index: $ \wh \a = \a i $ . 

The condition that $W_6$ be exact in terms of $K_5$, $W_6 = C_{0,2} X_4  = d K_5  $, implies the non-trivial conditions
\ba
 T_{( \wh \a \wh \b }{}^d K_{\wh \g \wh \d) d e f  }=  C_{( \wh \a \wh \b } X_{\wh \g \wh \d ) e f } 
\ea
\ba
\na_{(\wh \a} K_{\wh \b \wh \g ) d e f } - T_{( \wh \a \wh \b }{}^c K_{\wh \g ) c d e f }
= C_{( \wh a \wh \b} X_{\wh \g ) d e f} 
\ea
\ba
2 \na_{(\wh \a}  K_{\wh \b ) c d e f }  + 4 \na_{[c} K_{d e f ] \wh \a \wh \b }  + T_{\wh \a \wh \b }{}^b K_{b c d e f } 
= C_{\wh \a \wh \b} X_{c d e f} 
\ea
\ba
\na_{\wh \a } K_{b c d e f} - 5 \na_{[b} K_{c d e f ] \wh \a} = 0\ .  
\ea

\end{appendix}

\end{document}